\documentclass[12pt]{article}                 

\usepackage{latexsym,amsmath,amssymb,cite}      

\let\bra=\langle        \let\ket=\rangle 

\newcommand {\ud} {\mathrm{d}}
\newcommand {\cF} {{\cal F}}
\newcommand {\cA}{{\cal A}}

\newcommand {\cN}{{\cal N}}
\newcommand {\cO}{{\cal O}}

\newcommand {\cS}{{\cal S}}
\newcommand {\bbH}{\mathbb{H}}
\newcommand {\bbR}{\mathbb{R}}

\newcommand {\back}{\!\!\!\!\!\!\!\!\!\!\!\!}
\newcommand {\Back}{\!\!\!\!\!}
\newcommand {\im} {\mathrm{Im}}

\newcommand {\Li} {\mathrm{Li}_2}

\def\fnote#1#2{\begingroup\def\thefootnote{#1}\footnote{#2}
                \addtocounter{footnote}{-1}\endgroup}

\begin{document}                \baselineskip=16pt
\thispagestyle{empty}

\begin{flushright}
TUW--01/18\\
hep-th/0106159
\end{flushright}

\vspace*{1.5cm}

\begin{center}
{\LARGE 
{\bf Star Products from Open Strings \\[7pt]    in Curved Backgrounds
}}
\end{center}

\vspace*{5truemm}

\begin{center}
{\large Manfred Herbst,\fnote{\#}{e-mail: herbst@hep.itp.tuwien.ac.at}
        Alexander Kling,\fnote{$\,\Box$\,}{e-mail: kling@hep.itp.tuwien.ac.at}
        Maximilian Kreuzer\fnote{\,*\,}{e-mail: kreuzer@hep.itp.tuwien.ac.at}
}
\end{center}

{\sl 
\begin{center} 
Institut f\"ur Theoretische Physik, Technische Universit\"at Wien,\\
Wiedner Hauptstra\ss e 8-10, A-1040 Vienna, Austria
\end{center}
}

\vspace*{5mm}

\begin{abstract}        \normalsize

We define a non-commutative product for arbitrary gauge and B-field
backgrounds in terms of correlation functions of open strings. 
While off-shell correlations are, of course, not conformally invariant, 
it turns out that, at least to first derivative 
order, our product has the trace property and is associative up to surface 
terms if the background fields are put on-shell. No on-shell 
conditions for the inserted functions are needed, but it is essential to
include the full contribution of the Born-Infeld measure. We work with a 
derivative expansion and avoid any 
topological limit, which would effectively constrain $H$. 

\end{abstract}
\vfill


\clearpage
\setcounter{page}{1}

\section{Introduction and summary}
\label{sec:intro}

Noncommutative geometry as it turned out to arise from open strings in a 
background $B$-field~\cite{Connes:1998cr,Douglas:1998fm,Ardalan:1999ce,
Chu:1999qz,Schomerus:1999ug,Ardalan:2000av,Chu:2000gi,Seiberg:1999vs} has 
attracted much attention. 
Most of this and subsequent work was done in the context 
of a constant background. In terms of D-brane physics this corresponds to 
an embedding of a flat brane into a flat background. 
It is well known that in this situation the implications of the $B$ field 
background can effectively be described by replacing the ordinary 
(commutative) product of functions on the world volume of the D-brane 
by a noncommutative product, the Moyal product. 

There have been several attempts to generalize this picture to the 
situation of a non-constant background. This lead to a physical 
interpretation of Kontsevich's formula~\cite{Kontsevich:1997vb}, originally 
derived in the context of deformation quantization of the algebra of 
functions on Poisson manifolds,
in terms of the 
perturbative expansion of the path integral of a topological model 
of bosonic open strings~\cite{Cattaneo:2000fm}. A typical example of a 
Poisson manifold is provided by a symplectic manifold, i.e., a differentiable 
manifold naturally equipped with a closed non degenerate two form, 
$d\omega =0$. 
In terms of string theory this is related to a background $B$-field with 
vanishing field strength, $H=d B=0$. 

In this paper we address the problem of open strings in general backgrounds, 
in particular $B$ field backgrounds with nonvanishing field 
strength~\cite{Ho:2000fv,Cornalba:2001sm,Ho:2001qk}. In the
terminology of D-branes this is related to the embedding of curved branes 
into curved backgrounds. It has been argued that in this case the 
algebra of functions on the D-brane world volume becomes non-associative, 
with the non-associativity controlled by the field strength 
$H$~\cite{Cornalba:2001sm}. Nevertheless, it turned out that the product is 
still described in terms of Kontsevich's formula. 

Following a similar strategy as the authors of ref. \cite{Cornalba:2001sm}
we will expand about background fields to
extract 
the star product from correlation 
functions computed on the disk. We will work with a derivative expansion 
and will nowhere use a zero slope limit in our arguments.
Furthermore, we do not choose any gauge conditions for the 
background gauge fields. Here our setting deviates vitally from the one 
used in~\cite{Cornalba:2001sm}, where radial gauge was imposed on the two 
form gauge potential $B$. 
This choice of gauge allows them to extract the
part of the nonassociative star product deformation due to 
$H$, while the part due to $F=dA$ is given by Kontsevich's formula. 
Instead we prefer to work in a manifestly gauge invariant way. 
Furthermore, we only perform a perturbation expansion around 
the constant zero modes, and not an additional expansion around constant 
backgrounds, as done in~\cite{Cornalba:2001sm}. This keeps the full zero 
mode dependence of the background fields and simplifies the calculations.

Our main concern in this paper will be to discuss the 
properties of the product obtained by the procedure described above. 
Although this product is noncommutative and even nonassociative, we will 
show that associativity of the product of three functions and the trace
property for the integrated product of an arbitrary number of functions
hold, at least to first order in the derivative expansion,
up to surface 
terms. This is achieved by including the full Born-Infeld measure and  
the equations of motion of the spacetime background fields. However, 
no on-shell condition is needed for the functions inserted in the product! 
We find this result remarkable in view of a possible application of our
product for the description of correlation functions. Some work in this 
direction is in progress. 

Finally, we comment on the relation to the recent work of Cornalba and
Schiappa~\cite{Cornalba:2001sm}. Using the topological limit 
$g_{\mu\nu}\rightarrow 0$ 
they found that with the choice of radial gauge it is possible to adjust the 
integration measure in such a way that the integral still acts as a trace.
We rather use the measure that arises from the string theory correlations.
In this approach it turns out that the trace property of the integral is 
maintained when the background fields are put on-shell. This holds independent
of the gauge and even away from the topological limit.

The paper is organized as follows.
In section \ref{sec:model} we introduce the setup for the models
under consideration. We give the derivative expansions of 
the background fields in terms of Riemannian normal coordinates and 
introduce the additional interaction vertices. In section \ref{sec:prop} 
we review the calculations of \cite{Schomerus:1999ug} for the free field 
theory defined by the constant parts of the background 
fields and identify the effective open string parameters $G$ and $\Theta$. 
The vacuum amplitude of the free theory on the disk is computed in section 
\ref{sec:vacloops}. It contributes the ``Born-Infeld'' measure to the 
integration over the zero modes in the path integral. 
The relevant disk correlators are then presented in section \ref{sec:corr},
with some technical details given in the appendix. 
In section \ref{sec:properties} we extract 
a noncommutative and nonassociative Kontsevich-type product from these
correlators and discuss its properties. 
In particular we show that the trace property of the two-point function 
holds due to the equations of motion of the background fields. 
The ``Born-Infeld'' measure exactly cancels the additional contributions 
arising from partial integration. By the same mechanism the product of 
three functions does not depend on the way one introduces brackets, i.e. 
the nonassociativity is a surface term. This, in turn, implies the trace 
property for an arbitrary number of functions. We finish this section with 
some comments on the relations of our approach to the recent work of 
Cornalba and Schiappa. In particular we examine the implications of the 
radial gauge and the consistency of the topological limit used 
in~\cite{Cornalba:2001sm}. In the last section we
conclude with comments on some open questions and an outlook on further work.

{\bf Note added:} In the published version of this paper we argued
after equation (\ref{eq:2ordKonts}) that
an introduction of covariant derivatives in the product is justified,
because we calculated in Riemannian normal coordinates. Inclusion of the
full metric dependence in first derivative order shows, however, that
the partial derivatives in the product remain unchanged. In the
present version we correct
this statement, which affects equations (\ref{eq:derFterm}),
(\ref{eq:derivGTheta}) and the product (\ref{eq:2ordKonts}), as well
as related expressions thereafter. 

\section{The open string sigma model}
\label{sec:model}

The starting point of our considerations is the nonlinear sigma model of 
the bosonic open string\cite{Fradkin:1985qd,Abouelsaood:1987gd,Callan:1987bc}
\begin{eqnarray}
  \label{eq:sigmamodel}
  S &=& \frac 1{4\pi\alpha'} \int _\Sigma \,\ud^2\sigma \sqrt{h}
        {\Bigl(
         h^{ab} \partial_aX^\mu \partial_bX^\nu g_{\mu\nu}(X)
            +i \epsilon^{ab} \partial_aX^\mu \partial_bX^\nu B_{\mu\nu}(X)
         \Bigr)}\nonumber\\
    &+&  i \int _{\partial\Sigma} \,\ud s
         {\Bigl(
          \partial_s X^\mu A_\mu(X)
         \Bigr)},
\end{eqnarray}
which includes the spacetime metric $g_{\mu\nu}(X)$, the 2-form
gauge potential $B_{\mu\nu}(X)$ and the 1-form gauge potential $A_\mu(X)$.
$h_{ab}$ denotes the Euclidean metric on the world sheet $\Sigma$ and
$\ud s$ is the induced line element on the boundary.

In (\ref{eq:sigmamodel}) the boundary term with 
the 1-form gauge potential $A$ can be rewritten as a bulk term
\begin{equation}
  \label{eq:fieldstrength}
  \int _\Sigma \,\ud^2\sigma \sqrt{h} \,
        {i \epsilon^{ab} \partial_aX^\mu \partial_bX^\nu F_{\mu\nu}(X)},
\end{equation}
where $F = \ud A$ is the corresponding 2-form field strength.

Both, the 1-form potential $A$ and the 2-form potential $B$, are associated 
with spacetime gauge invariances. For the former the gauge transformation
\begin{equation}
  \label{eq:Agaugetrans}
  \delta A = \ud \lambda
\end{equation}
leaves the action (\ref{eq:sigmamodel}) invariant. In open string
theory there does not exist a gauge transformation for the 2-form 
potential $B$ alone, because surface terms require a combined transformation
\begin{eqnarray}
  \label{eq:BAgaugetrans}
  \delta B &=& \ud \Lambda , \nonumber \\
  \delta A &=& - \frac \Lambda{2\pi\alpha'}
\end{eqnarray}
that does not change the action (\ref{eq:sigmamodel}). 
{}From (\ref{eq:Agaugetrans}) and (\ref{eq:BAgaugetrans}) one can see that 
the combination $\cF = B + 2\pi\alpha' F = B + 2\pi\alpha' \ud A$ is
invariant under both gauge symmetries. Therefore, gauge invariant expressions
contain the 2-form $\cF$ and the 3-form field strength
$H = \ud \cF = \ud B$.

If one considers a brane that is not spacetime 
filling, the gauge field $A$ and hence $\cF$, as well as the resulting
noncommutative geometry, are only defined on that 
brane.
Furthermore, 
in topologically nontrivial backgrounds, the gauge potentials $A$ and $B$ 
may not be globally 
defined. These issues 
are, however, irrelevant in the present context.

In the classical approximation of open string theory the world sheet 
$\Sigma$ is a disk. Taking advantage of the conformal invariance 
of the theory, we map the disk to the upper half plane $\bbH$ and perform
our calculations there. Furthermore, we choose the conformal gauge
and use complex coordinates $z = \sigma^1 + i \sigma^2$. Thus the 
world sheet metric becomes 
$h_{z\bar z} = e^{2\omega(z,\bar z)} \delta_{z\bar z}$ and 
the invariant line element at the boundary is 
$\ud s = e^\omega \ud \tau$. The derivatives tangential and 
normal to the boundary are $\partial_\tau = (\partial + \bar \partial)$ and 
$\partial_n = i(\bar \partial - \partial)$, respectively. 
In this parametrization the action (\ref{eq:sigmamodel}) 
is given by 
\begin{eqnarray}
  \label{eq:actionconf}
  S = \frac 1{2\pi\alpha'} \int _\bbH \, \ud^2z \,
      {\partial X^\mu \bar\partial X^\nu 
       \Bigl(g_{\mu\nu}(X) + \cF_{\mu\nu}(X)\Bigr)},
\end{eqnarray}
and the corresponding mixed boundary condition along the brane is
\begin{eqnarray}
  \label{eq:boundary}
   g_{\mu\nu}(X)  (\partial-\bar\partial) X^\nu - 
   \cF_{\mu\nu}(X) (\partial+\bar\partial) X^\nu \biggr|_{\bar z = z}= 0.
\end{eqnarray}

Following the standard procedure we expand the field 
$X^\mu(z,\bar z)$ around the constant zero mode contribution 
$x$~\cite{Fradkin:1985qd},
\begin{equation}
  \label{eq:quantumfield}
  X^\mu(z,\bar z) = x^\mu + \zeta^\mu(z,\bar z),
\end{equation}
so that the path integral over the field $X^\mu(z,\bar z)$ splits into an
ordinary integral over the constant zero modes $x^\mu$ and a
path integral over the quantum fluctuations $\zeta^\mu(z,\bar z)$
\begin{eqnarray}
  \label{eq:pathint}
  \bra \,:\!f_1[X(z_1)]\!: &\ldots& :\!f_N[X(z_N)]\!: \,\ket 
  \quad = \nonumber \\
  &=& \int [\ud X]\, e^{-S[X]} f_1[X_1]\! \ldots f_N[X_N] 
  \quad = \nonumber\\
  &=& \int \ud^D x \int [\ud \zeta]\, e^{-S[x\!+\!\zeta]}
  f_1[x\!+\!\zeta_1] \ldots f_N[x\!+\!\zeta_N]\; ,
\end{eqnarray}
where the functions $f_i[X(z_i)]$ denote arbitrary
insertions into the path integral. 
Before we expand
the action $S[X] = S[x + \zeta]$ around the zero modes we simplify our
computation by choosing Riemannian normal coordinates
\cite{Alvarez-Gaume:1981hn,Braaten:1985is},
\begin{eqnarray}
  \label{eq:Riemannexpand}
  g_{\mu\nu}(x+\zeta) &=& \eta_{\mu\nu} - 
                    \frac 13 R_{\mu\rho\nu\sigma}(x) \zeta^\rho \zeta^\sigma + 
                    \cO(\zeta^3) , \\
  \label{eq:Fexpand}
  \cF_{\mu\nu}(x+\zeta) &=& \cF_{\mu\nu}(x)+ 
                      \partial_\rho \cF_{\mu\nu}(x) \zeta^\rho +
                      \frac 12 \partial_\rho \partial_\sigma 
                      \cF_{\mu\nu}(x) \zeta^\rho \zeta^\sigma +
                      \cO(\zeta^3).
\end{eqnarray}
In contrast to~\cite{Cornalba:2001sm} we do not choose 
radial gauge for $\cF_{\mu\nu}(X)$. In that case (\ref{eq:Fexpand})
would split into two separate expansions for $B$ and $F$, where 
the non-constant part of the $B$ expansion contains only the field 
strength $H$. 
With (\ref{eq:Riemannexpand}) and (\ref{eq:Fexpand}) we are able to 
write the action (\ref{eq:actionconf}) as
\begin{eqnarray}
  \label{eq:pertexp}
  S = \frac 1{2\pi\alpha'} \int _\bbH \ud^2z \Back & & \Back
        {\Bigl\{\partial \zeta^\mu \bar\partial \zeta^\nu 
         (\eta_{\mu\nu} + \cF_{\mu\nu})} +
         \partial \zeta^\mu \bar\partial \zeta^\nu \zeta^\rho \,
         \partial_\rho \cF_{\mu\nu} + \nonumber\\
     &+& \partial \zeta^\mu \bar\partial \zeta^\nu \zeta^\rho \zeta^\sigma \,
         ( \frac 12 \partial_\rho \partial_\sigma \cF_{\mu\nu}
         - \frac 13 R_{\mu\rho\nu\sigma} )
         + \cO(\partial_{\rho}^3) \Bigr\}.
\end{eqnarray}

In the following we will restrict our considerations to terms of 
at most first order in derivatives of the spacetime background fields.

\section{The free propagator}
\label{sec:prop}

As a warm up for later calculations and to set up the relevant 
techniques of our approach let us first calculate the propagator for the 
free field theory defined by the Gaussian part of (\ref{eq:pertexp})
in the path integral,
\begin{equation}
  \label{eq:freeaction}
  S_\mathrm{free} = \frac 1{2\pi\alpha'} \int _\bbH \, \ud^2z \,
        {\partial \zeta^\mu \bar\partial \zeta^\nu \eta_{\mu\nu}}
    + \frac i{4\pi\alpha'} \oint _{\partial\bbH} \, \ud \tau \,
      {\zeta^\mu \partial_\tau \zeta^\nu \cF_{\mu\nu}}.
\end{equation}
Here $\partial\bbH$ denotes the boundary of the upper half plane, i.e., 
the real line.\footnote{We have used the divergence theorem for complex 
coordinates, which reads 
$\int _\Sigma\,\ud^2z \,{(\partial_z v^z \pm\partial_{\bar z}v^{\bar z})}=i\oint _{\partial \Sigma} \,{(\ud \bar z v^z \mp \ud z v^{\bar z} )}.$}
The second term contributes to the boundary condition which 
takes the same form as (\ref{eq:boundary}) with $\eta_{\mu\nu}$ and 
$\cF_{\mu\nu}(x)$ replacing the full metric $g_{\mu\nu}(X)$ and 
$\cF_{\mu\nu}(X)$, respectively. The boundary term can be regarded as a 
perturbative correction~\cite{Schomerus:1999ug} to the free propagator
\begin{equation}
  \label{eq:freeprop}
  \bra\zeta^\mu(u,\bar u)\,\zeta^\nu(w,\bar w)\ket = 
                      - \frac {\alpha'}2 \eta^{\mu\nu}\ln |u-w|^2
                      - \frac {\alpha'}2 \eta^{\mu\nu}\ln |u-\bar w|^2.
\end{equation}
The homogeneous (image charge) part accounts for the Neumann boundary condition
$\partial_n \zeta^\mu |_{\partial \bbH} = 0$ of the theory without 
perturbation. The propagator of the perturbed theory is then given by
the (connected) 2-point correlation
\begin{equation}
  \label{eq:freepropF}
  \bra \zeta^\mu(u,\bar u)\,\zeta^\nu(w,\bar w)\ket_\cF = 
  \bra \zeta^\mu(u,\bar u)\,\zeta^\nu(w,\bar w) 
      \mathrm{e}^{-\frac i{4\pi\alpha'} 
                 \oint _{\partial\bbH} \, \ud \tau \,
                 {\zeta^\rho \partial_\tau \zeta^\sigma \cF_{\rho\sigma}}}\ket.
\end{equation}
Disconnected loop contributions will only contribute to the measure (see
below).
Expanding in a perturbation series the term of order n 
\begin{equation}
  \label{eq:ninsertion}
\bra\zeta^\mu(u,\bar u)\zeta^\nu(w,\bar w)\frac 1{n!}
                  \Bigl\{\frac i{4\pi\alpha'} 
                  \Bigl[\oint _{\partial\bbH} \, \ud z \,
                  {\partial \zeta^\rho \zeta^\sigma \cF_{\rho\sigma}} +
                  \oint _{\partial\bbH} \, \ud \bar z \,
                  {\bar \partial \zeta^\rho \zeta^\sigma \cF_{\rho\sigma}}
                  \Bigr] \Bigr\}^n \ket
\end{equation}
gives two slightly different contributions, depending on whether n is 
even or odd. By using the derivative of the propagator 
(\ref{eq:freeprop}) it is straightforward to obtain the 
result\footnote{In this calculation there appear integrals of the form
$\oint _{\partial \bbH} \, \ud z \frac1{\bar u - z} \frac1{\bar z - w}$.
The part along the real axis $\bbR$ is 
$\int _\bbR \, \ud r \frac1{\bar u - r} \frac1{r - w}$, whereas the integral
along the semicircle in the upper half plane with infinite radius is zero.
Therefore, the original integral can be written as
\begin{displaymath}  
  \oint _{\partial \bbH} \, \ud z \frac1{\bar u - z} \frac1{\bar z - w} = 
  \oint _{\partial \bbH} \, \ud z \frac1{\bar u - z} \frac1{z - w},
\end{displaymath}
which can be evaluated using the residue theorem.}
\begin{eqnarray}
  \label{eq:ncontribution}
  \frac i{2\pi} (\cF^n)_{\lambda\kappa}
  \Bigl\{(-1)^{n-1} 
  \oint _{\partial \bbH} \,\ud z \, {\eta^{\mu\lambda} \frac1{\bar u-z}
   \bra \zeta^\kappa \zeta^\nu(w,\bar w)\ket} \,\,\, & & \nonumber\\
  + \oint _{\partial \bbH} \,\ud \bar z \,{\eta^{\mu\lambda} \frac1{u-\bar z}
   \bra \zeta^\kappa \zeta^\nu(w, \bar w)\ket}
  \Bigr\} &.&
\end{eqnarray}
The remaining divergent integrals are regularized by differentiating with 
respect to $w$ and $\bar w$, respectively. This yields a finite result plus 
an infinite additive constant $C^{\mu\nu}_{(\infty)}$,
\begin{equation}
  \label{eq:regresult}
  \alpha' (\cF^n)^{\mu\nu} 
  \Bigl\{(-1)^{n-1} \ln(\bar u-w) - \ln(u-\bar w) \Bigr\} + 
  C^{\mu\nu}_{(\infty)}.
\end{equation}
Now it is possible to sum up all orders in a geometrical series, which 
finally gives the desired propagator~\cite{Abouelsaood:1987gd,Callan:1987bc}
\begin{eqnarray}
  \label{eq:fullprop}
  \bra \zeta^\mu(u,\bar u)\,\zeta^\nu(w,\bar w)\ket_\cF 
  \hspace{-3pt}&\hspace{-5pt}=\hspace{-5pt}&\hspace{-3pt} 
  -\alpha' \Bigl\{\eta^{\mu\nu} (\ln |u-w| - \ln |u-\bar w|) \nonumber\\
 \hspace{-3pt}&\hspace{-7pt}+ \hspace{-7pt}&\hspace{-3pt}
   G^{\mu\nu} \ln |u-\bar w|^2 
  -\Theta^{\mu\nu} \ln \Bigl(\frac {\bar w - u}{\bar u-w}\Bigr)\Bigr\} 
  + C^{\mu\nu}_{(\infty)},
\end{eqnarray}
where we have introduced the quantities\footnote{For later reference 
        we have expressed $G^{\mu\nu}$ and $\Theta^{\mu\nu}$ using the full
        bulk metric $g_{\mu\nu}$, whereas the terms in 
        (\ref{eq:fullprop}) contain, of course, the Minkowski metric 
        $\eta_{\mu\nu}$ because of the
        Riemannian normal coordinates.}
\begin{equation}
  \label{eq:openmetric}
  G^{\mu\nu} := \Bigl(\frac 1{g-\cF}~g~\frac 1{g+\cF} \Bigr)^{\mu\nu} 
  \quad \mathrm{and} \quad
  \Theta^{\mu\nu} := -\Bigl(\frac 1{g-\cF}~\cF~\frac 1{g+\cF} \Bigr)^{\mu\nu}.
\end{equation}
The integration constant $C^{\mu\nu}_{(\infty)}$ plays no 
essential role and can be set to a convenient value, e.g. 
$C^{\mu\nu}_{(\infty)}=0$~\cite{Seiberg:1999vs}. 
When restricted to the boundary ($u = \bar u = \tau$
and $w = \bar w = \tau'$) the propagator reduces to the simple form
\begin{eqnarray}
  \label{eq:boundaryprop}
  \alpha' i \pi \Delta^{\mu\nu}(\tau,\tau') 
  &:=& \bra \zeta^\mu(\tau)\,\zeta^\nu(\tau')\ket_\cF \nonumber\\
  & =& -\alpha' G^{\mu\nu} \ln (\tau-\tau')^2
       -\alpha' i \pi \Theta^{\mu\nu} \epsilon(\tau-\tau').
\end{eqnarray}
As discussed in~\cite{Seiberg:1999vs} the boundary 
propagator (\ref{eq:boundaryprop}) suggests to interpret $G_{\mu\nu}$ as an
effective metric seen by the open strings, in contrast to $g_{\mu\nu}$, which 
is to be viewed as the closed string metric in the bulk. 

For later purposes we elaborate on the distinction between the open 
string quantities $G^{\mu\nu}$ and $\Theta^{\mu\nu}$ and the closed 
string quantities $g_{\mu\nu}$ and $B_{\mu\nu}$. 
In order to make a clear distinction between the bulk and the boundary 
quantities, we mark all expressions that refer to boundary quantities 
with bars. To this end we define
\begin{eqnarray}
  \label{eq:boundquant}
  \bar G_{\mu\nu} := (g-\cF^2)_{\mu\nu} \quad \mathrm{and} \quad
  \bar \Theta^\mu{}_\nu := -\cF^\mu{}_\nu.
\end{eqnarray}
The first of the above  definitions is equivalent to setting 
$\bar G^{\mu\nu} = G^{\mu\nu}$ and requiring $\bar G_{\mu\nu}$ to be its 
inverse. The second definition follows from setting 
$\bar\Theta^{\mu\nu}=\Theta^{\mu\nu}$ and pulling indices with 
$\bar G_{\mu\nu}$. In an analogous way we label all expressions that are
built out of these quantities with bars, e.g. the Christoffel symbol 
$\bar \Gamma_{\mu\nu}{}^\rho$ and the covariant derivative 
$\bar D_\mu$ compatible with the open string metric $\bar G_{\mu\nu}$. 

\section{Vacuum amplitude and integration measure}
\label{sec:vacloops}

Let us now consider loop 
contributions arising from an even number of insertions of the boundary 
perturbation of (\ref{eq:freeaction})\footnote{Odd powers vanish because of 
the antisymmetry of $\cF_{\mu\nu}$}. In this calculation there appear
divergences when the insertion points approach the boundary. We regularize
these terms by keeping a fixed distance $d_0$ with respect to the
metric in conformal gauge to the boundary $\partial \bbH$, i.e., we impose
$|z - \bar z| \geq 2\im (z) \ge e^{-\omega} d_0$. To make this more explicit 
let us consider the one loop contribution of the $\cF^2$ term,
\begin{eqnarray}
  \label{eq:F2loop}
  \frac 12 
    \bra \Bigl( \frac {-i}{4\pi\alpha'} \Bigr)^2
      \oint _{\partial\bbH} \, \ud \tau \,
      {\zeta^\mu \partial_\tau \zeta^\nu \cF_{\mu\nu}} \times
      \oint _{\partial\bbH} \, \ud \tau' \,
      {\zeta^\rho \partial_\tau' \zeta^\sigma \cF_{\rho\sigma}} 
      \ket_\mathrm{1-loop}.
\end{eqnarray}
Using the same techniques as for the chains (\ref{eq:ninsertion}) gives the
divergent contribution
\begin{eqnarray}
  \label{eq:F2loopresult}
  \Bigl( \frac 1{4\pi d_0} \int \ud s \Bigr) 
  \frac 12 \cF_\mu{}^\nu \cF_\nu{}^\mu,
\end{eqnarray}
where $\ud s = \ud\tau e^\omega$ is the invariant line element in
conformal gauge. Summing up all powers of $\cF$ in the 1-loop 
contribution yields
\begin{eqnarray}
  \label{eq:allFloop}
  \Bigl(\frac 1{4\pi d_0} \int \ud s\Bigr) 
  \sum_{n=1}^\infty \frac 1{2n} \cF^{2n} = 
  - \Bigl(\frac 1{4\pi d_0} \int \ud s\Bigr) 
  \frac 12 \ln (\det (\delta - \cF^2)_\mu{}^\nu).  
\end{eqnarray}
As observed in~\cite{Tseytlin:1986zz,Tseytlin:2000mt} this linear divergence 
is in fact regularization scheme dependent and can be absorbed into the 
tachyon by a field redefinition. But a finite constant part
\begin{eqnarray}
  \label{eq:allFfinit}
  b_0 \ln (\det (\delta - \cF^2)_\mu{}^\nu)
\end{eqnarray}
may remain after subtraction of appropriate counter terms. The
analysis given in~\cite{Fradkin:1985qd,Tseytlin:2000mt} 
determined the constant
$b_0$ to be $\frac 14$ in order to yield the Born-Infeld action
for a vanishing tachyon field.

In (\ref{eq:allFfinit}) we have added up all powers of $\cF$ contributing to 
the connected vacuum graphs. Taking into account all disconnected one loop 
graphs to all orders of the interaction leads to the 
Born-Infeld Lagrangian
\begin{eqnarray}
  \label{eq:measureterm}
  \sum_{n=0}^\infty \frac 1{n!} 
  \bigl(\ln (\det (\delta - \cF^2)_\mu{}^\nu)^{\frac 14}\bigr)^n = 
  \root 4 \of {\det (\delta - \cF^2)_\mu{}^\nu} =
  \sqrt {\det (\delta - \cF)_\mu{}^\nu}~.
\end{eqnarray}
Here we used the antisymmetry of $\cF_{\mu\nu}$ to change the sign in the 
determinant. Expression (\ref{eq:measureterm}) can also be interpreted as
a contribution to the measure of the integration
over the zero modes in the path integral. Although we make use of
Riemannian normal coordinates for the perturbation expansion, we can
write the measure in a covariant way by including the term 
$\sqrt {\det g_{\mu\nu}}$. Therefore, if there are no operator insertions 
in the path integral (\ref{eq:pathint}), we obtain the Born-Infeld action
\begin{eqnarray}
  \label{eq:BIAction}
  \int \ud^{D} x \sqrt {\det g_{\mu\nu}} 
                 \root 4 \of {\det (\delta - \cF^2)_\mu{}^\nu} &=&
  \int \ud^{D} x \root 4 \of {\det g_{\mu\nu}} 
                 \root 4 \of {\det \bar G_{\mu\nu}} = \nonumber \\
  &=& \int \ud^{D} x \sqrt{\det (g - \cF)_{\mu\nu}},
\end{eqnarray}
where $\bar G_{\mu\nu}$ is the boundary metric as defined in 
(\ref{eq:boundquant}).

So far we have taken into account all possible diagrams of the boundary
insertion of (\ref{eq:freeaction}).
Therefore, we can now work with the full propagator (\ref{eq:fullprop})
for all higher order interaction terms.
For the remainder of the paper we make use of the abbreviations
$g = \det g_{\mu\nu}$ and
$\int_x =\int\ud^{D}x\sqrt{g-\cF}=\int\ud^{D}x\root 4\of{g}\root 4\of{\bar G}$.
Furthermore, we set $2\pi\alpha'=1$.

\section{Correlation functions}
\label{sec:corr}

In string theory interactions of different particles of the string spectrum 
are calculated by inserting the corresponding vertex operators
into the path integral. Our goal is to extract a noncommutative
product of functions from the open string theory correlation 
functions~\cite{Cornalba:2001sm}. To this end we do not 
restrict ourselves to on-shell vertex operators, but investigate
the correlator of two general functions $f[X(\tau)]$ and $g[X(\tau')]$ 
allowed to be off-shell. To simplify the calculations we
take the order of insertions to be $\tau < \tau'$.
Since the functions are composed operators,
one has to introduce an appropriate normal ordering. 
As shown in appendix A.4 it is given by
\begin{eqnarray}
  \label{eq:normalorder}
 &&\hspace*{-15pt} : \zeta^\mu (\tau) \, \zeta^\nu (\tau') : \hspace*{6pt}= 
    \hspace*{5pt}\zeta^\mu (\tau) \, \zeta^\nu (\tau')\nonumber 
\\
 &&\hspace*{-15pt}  
    +\frac {1}{2\pi} G^{\mu\nu} \ln(\tau-\tau')^2    
     +\frac {1}{2\pi} 
     \partial_\rho G^{\mu \nu} \ln(\tau-\tau')^2
     \; \zeta^\rho (\frac {\tau+\tau'}{2}) .
\end{eqnarray}
Taking into account the subtractions (\ref{eq:normalorder}) the free
propagator (\ref{eq:boundaryprop}) yields~\cite{Schomerus:1999ug}
\begin{eqnarray}
  \label{eq:MoyalContr}
  & &\bra\,:\!f[X(\tau)]\!: \; :\!g[X(\tau')]\!: \,\ket_{\cF} 
   = \nonumber\\
  &=&\int _x \sum _{n=0}^\infty 
             \frac 1{n!} \Bigl(\frac i2 \Bigr)^n
             \Delta^{\mu_1\nu_1} \ldots \Delta^{\mu_n\nu_n} \;
             \partial_{\mu_1} \ldots \partial_{\mu_n} f(x) \;
             \partial_{\nu_1} \ldots \partial_{\nu_n}g(x) \\
  &=&\int _x \sum _{n=0}^\infty 
             \frac 1{n!} \Bigl(\frac{-1}{2\pi}\Bigr)^n
             G^{\mu_1\nu_1}\!..~G^{\mu_n\nu_n} \ln^n(\tau - \tau')^2\;
             \partial_{\mu_1}\!..~\partial_{\mu_n} f(x) \; * \;
             \partial_{\nu_1}\!..~\partial_{\nu_n}g(x) . \nonumber
\end{eqnarray}
In the last line we have summarized all $\Theta^{\mu\nu}$-dependent 
contributions in the product
\begin{eqnarray}
  \label{eq:MoyalPart}
  f * g &=& \sum _{n=0}^\infty 
             \frac 1{n!} \Bigl(\frac i2 \Bigr)^n
             \Theta^{\mu_1\nu_1}(x) \ldots \Theta^{\mu_n\nu_n}(x) \;
             \partial_{\mu_1} \ldots \partial_{\mu_n} f(x) \;
             \partial_{\nu_1} \ldots \partial_{\nu_n}g(x) \nonumber\\
        &=& e^{\frac i2 \Theta^{\mu\nu}(z) \partial_{x^\mu}\partial_{y^\nu}} 
            f(x) g(y)\Bigr\vert_{x=y=z} ,
\end{eqnarray}
which we will refer to as ``Moyal like'' part of the final
non-commutative product. It has the well known structure of the Moyal product
and reduces to it if $\Theta^{\mu\nu}$ is constant.  
In this case (\ref{eq:MoyalPart}) is clearly associative and satisfies the 
trace property. This is, however, no longer true, if $\Theta^{\mu\nu}$ is a 
generic field.

Going one step further in the derivative expansion we have to take into 
account the contribution to the non-commutative product arising from the 
interaction term%
\footnote{We reintroduce general coordinates in order to see the full
  dependence on the bulk metric $g$.
}%
\begin{eqnarray}
  \label{eq:derFterm}
  \frac 1{2\pi\alpha'} \int _\bbH \ud^2z \,
         \partial \zeta^\mu \bar\partial \zeta^\nu \zeta^\rho \,
         \partial_\rho (g+\cF)_{\mu\nu} .
\end{eqnarray}
The rather cumbersome calculations are explained in the appendix.
Using (\ref{eq:tifinite}) and (\ref{eq:tkfinite}) we obtain
\begin{eqnarray}
  \label{eq:dFcorr}
  \bra\;:\!f[X(\tau)]\!: \, :\!g[X(\tau')]\!:\;\ket_{\partial \cF}
      \; &=& \nonumber \\
  - \frac 1{12} \int _x 
              \Theta^{\mu\rho}\partial_\rho \Theta^{\nu\sigma} 
              \Back &\bigl(& \Back
                    \partial_\mu \partial_\nu f * \partial_\sigma g +
                    \partial_\sigma f * \partial_\mu \partial_\nu g
              \bigr) \\
  - \frac i{8\pi} \int _x 
              \Theta^{\mu\rho}\partial_\rho G^{\nu\sigma} 
              \Back &\bigl(& \Back
                    \partial_\nu \partial_\sigma f * \partial_\mu g -
                    \partial_\mu f * \partial_\nu \partial_\sigma g
              \bigr) \ln (\tau-\tau')^2 \nonumber \\
  - \frac i{4\pi} \int _x 
              G^{\nu\sigma}\partial_\sigma \Theta^{\rho\mu}
              \Back &\bigl(& \Back
                    \partial_\nu \partial_\rho f * \partial_\mu g -
                    \partial_\mu f * \partial_\nu \partial_\rho g
              \bigr) \ln (\tau-\tau')^2 \nonumber \\
  - \frac 1{16\pi^2} \int _x 
                \bigl(G^{\mu\rho}\partial_\rho G^{\nu\sigma} \!\!-
                      2G^{\nu\rho}\partial_\rho G^{\sigma\mu}
                \bigl)
              \Back &\bigl(& \Back
                    \partial_\nu \partial_\sigma f * \partial_\mu g +
                    \partial_\mu f * \partial_\nu \partial_\sigma g
              \bigr) \ln^2(\tau-\tau')^2 +\ldots, \nonumber
\end{eqnarray}
where we only
kept the $\Theta^{\mu\nu}$ terms from the contributions of the free 
propagator (\ref{eq:fullprop}), since the $G^{\mu\nu}$ parts are
irrelevant for our further discussion, as we shall see shortly.
Only the first line of (\ref{eq:dFcorr}) will contribute to our 
non-commutative product. The partial derivatives of the fields imply that
the whole expression (\ref{eq:dFcorr}) vanishes for constant background fields.

We define now the non-commutative product as
\begin{eqnarray}
  \label{eq:defKonts}
  \sqrt {g-\cF} \; f(x) \; \circ \; g(x)
  &:=& \int [\ud \zeta]\, e^{-S[x\!+\!\zeta]}
  f[X(0)] \; g[X(1)] \; .
\end{eqnarray}
The choice of the distance $\tau'-\tau = 1$ is such that the scale
dependent contributions of (\ref{eq:MoyalContr}) and (\ref{eq:dFcorr})
are removed.%
\footnote{%
	The value 1 is due to our choice of the infrared cut-off, i.e., 
	the constant $C^{\mu\nu}_{(\infty)}$ in (\ref{eq:fullprop}).}
The resulting non-commutative product is the scale and translation invariant
part of the 2-point correlation. This product is independent of $G^{\mu\nu}$, 
and we 
will see that only this part of the correlation has appropriate off-shell 
properties
(as long as the background fields are on-shell). The full off-shell 
correlations will, of course, also have $G^{\mu\nu}$-dependent contributions.


{}From (\ref{eq:MoyalContr}) and (\ref{eq:dFcorr}) we see that, up to first 
order in derivatives of $\Theta^{\mu\nu}$, the product 
reads\footnote{Subsequently we abbreviate 
$\cO\bigl(\partial\Theta)^2,\partial^2\Theta\bigr)$ by $\cO(\partial^2)$.}
\begin{eqnarray}
  \label{eq:2ordKonts}
  f(x) \; \circ \; g(x) \;=\; f * g 
  \Back&-&\Back \frac 1{12}
              \Theta^{\mu\rho} \partial_\rho \Theta^{\nu\sigma} \;
              \Bigl(\partial_\mu \partial_\nu f \;* \partial_\sigma g +
                    \partial_\sigma f \;* \partial_\mu \partial_\nu g \Bigr) +\nonumber\\
  &+& \cO\bigl((\partial\Theta)^2,\partial^2\Theta\bigr) .
\end{eqnarray}

A comparison of (\ref{eq:2ordKonts}) with the formula given in
\cite{Kontsevich:1997vb}
shows that
the non-commutative
product (\ref{eq:defKonts}) coincides with the Kontsevich formula.
We do not require, however, that the field $\Theta^{\mu\nu}$ defines
a Poisson structure. 

\section{Properties of the non-commutative product}
\label{sec:properties}

In the limit $\alpha' \!\rightarrow\! 0$ the correlator of an arbitrary
number of functions in the presence of a closed $B$-field background can 
be evaluated by integration of the non-commutative product of these 
functions. On the disk the $SL(2,\bbR)$ invariance of the correlators 
requires the product to satisfy the trace property.

The non-commutative product (\ref{eq:defKonts}) defined without
the use of this limit, however, 
does not describe the full correlation functions,
because the $G^{\mu\nu}$-dependent contractions give additional contributions. 
Even so, we will show in this section that the trace property can be 
maintained for the product (\ref{eq:2ordKonts}) if one imposes the 
equations of motion for the background fields, while the inserted 
functions are allowed to stay completely generic.

In string theory the background field equations of motion are related
to the renormalization group $\beta$ functions which probe the breaking
of Weyl invariance (and hence the conformal invariance) of the theory.
Since we perform our calculations up to first order in derivatives
of the background fields, we expect that we only have to account for the
generalization of the Maxwell equation~\cite{Dorn:1986jf,Callan:1987bc},
\begin{eqnarray}
  \label{eq:eom}
  G^{\rho\sigma}D_\rho \cF_{\sigma\mu}
  -\frac 12 \Theta^{\rho\sigma}H_{\rho\sigma\lambda} \cF^\lambda{}_\mu = 0 .
\end{eqnarray}
To show our proposition we rewrite
(\ref{eq:eom}) in a more appropriate way,
\begin{eqnarray}
  \label{eq:eomTheta}
  \partial_\mu \Bigl(\sqrt{g-\cF}~\Theta^{\mu\nu} \Bigr) \!&=&\!
  \sqrt {g} D_\mu \Bigl(
  \frac {\root 4 \of {\bar G}}{\root 4 \of g}~\Theta^{\mu\nu} \Bigr)  =
\\
  \!&=&\! - \sqrt{g-\cF}
  \Bigl(G^{\rho\sigma}D_\rho \cF_{\sigma\mu}
  -\frac 12 \Theta^{\rho\sigma}H_{\rho\sigma\lambda} \cF^\lambda{}_\mu\Bigr)
  G^{\mu\nu} = 0~, \nonumber
\end{eqnarray}
where we have used the relation
\begin{eqnarray}
  \label{eq:relGamma}
  \Theta^{\rho\sigma} D_\mu \cF_{\rho\sigma} = 
  - \frac 12 \bar G^{\rho\sigma} D_\mu \bar G_{\rho\sigma} =
  \Gamma_{\mu\lambda}{}^\lambda - \bar \Gamma_{\mu\lambda}{}^\lambda ,
\end{eqnarray}
and the fact that the quotient
$\frac {\root 4 \of {\bar G}}{\root 4 \of g}$ is a scalar. We introduce
the usual notation~$\approx$~for equivalence up to equations of motion. 
Note, furthermore, that in the following all relations are valid only
to first order in derivatives of $\Theta^{\mu\nu}$.

We start with the product of two functions and show that (\ref{eq:2ordKonts})
is symmetric under the integral
\begin{eqnarray}
  \label{eq:trace2}
  \int\ud^{D} x \sqrt {g-\cF} \; f \; \circ \; g &\approx& 
  \int\ud^{D} x \sqrt {g-\cF} \; g \; \circ \; f .
\end{eqnarray}
This relation holds due to (\ref{eq:eom}) and (\ref{eq:eomTheta}), 
because then the first order term in $\Theta^{\mu\nu}$
of (\ref{eq:2ordKonts}) becomes a total divergence,
\begin{eqnarray}
  \label{eq:totaldiv}
  \int \ud^{D} x \sqrt {g-\cF} \; \Theta^{\mu\nu} \; \partial_\mu f \; \partial_\nu g
  \approx \int \ud^{D} x \; \partial_\mu \Bigl(
    \sqrt {g-\cF} \; \Theta^{\mu\nu} \; f \; \partial_\nu g \Bigr) = 0,
\end{eqnarray}
and the remaining antisymmetric parts can be written as contributions
of second order in derivatives. Notice that here and in the subsequent 
relations it is essential that the constant $b_0$ in the integration 
measure takes the value $\frac 14$ in order to produce the total divergence.

For a general field $\Theta^{\mu\nu}$ the product (\ref{eq:2ordKonts})
is not associative. But, applying (\ref{eq:eom},\ref{eq:eomTheta})  again,
associativity up to surface terms is ensured for the product of three 
functions. To see this we calculate $(f\circ g)\circ h-f\circ (g\circ h)$.
Using the formula 
\begin{eqnarray}
  \label{eq:diffMoyal}
  \partial_\rho \bigl(f*g\bigr)(x) &=& 
  (\partial_{x^\rho} + \partial_{y^\rho} + \partial_{z^\rho})
  e^{\frac i2 \Theta^{\mu\nu}(z) \partial_{x^\mu}\partial_{y^\nu}} 
            f(x) g(y)\Bigr\vert_{x=y=z} \nonumber\\
  &=& \partial_\rho f * g + f * \partial_\rho g + 
  \frac i2 \partial_\rho \Theta^{\mu\nu}\partial_\mu f * \partial_\nu g.
\end{eqnarray}
for the product (\ref{eq:MoyalPart}) we obtain
\begin{eqnarray}
  \label{starnonassoc}
  (f * g) * h &=& [f*g*h] + 
                \frac 14 \Theta^{\mu\sigma} \partial_\sigma \Theta^{\nu\rho}
                \partial_\nu f * \partial_\rho g * \partial_\mu h + \cO(\partial^2),\nonumber\\
  f * (g * h) &=& [f*g*h] - 
                \frac 14 \Theta^{\mu\sigma} \partial_\sigma \Theta^{\nu\rho}
                \partial_\mu f * \partial_\nu g * \partial_\rho h + \cO(\partial^2),
\end{eqnarray}
where $[f*g*h]$ denotes the part with no derivatives acting on 
$\Theta^{\mu\nu}$. So the non-associativity reads
\begin{eqnarray}
  \label{eq:nonassoc}
  (f \circ g) \circ h \Back&-&\Back f \circ (g \circ h) \; = \nonumber\\
  &=& \frac{1}{6} \bigl( \Theta^{\mu\sigma}  \partial_\sigma \Theta^{\nu\rho} +  
     (\textrm{cycl.}{}^{\mu\nu\rho}) \bigr) \, 
     \partial_\mu f * \partial_\nu g * \partial_\rho h + \cO(\partial^2) \nonumber\\
  &=& \frac{1}{6} 
     \Theta^{\mu\sigma} \Theta^{\nu\lambda} \Theta^{\rho\kappa}
     \bar H_{\sigma\lambda\kappa} \, 
     \partial_\mu f * \partial_\nu g * \partial_\rho h + \cO(\partial^2) .
\end{eqnarray}
In the last line we have introduced the 3-form field
$\bar H = \ud (\Theta^{-1})$, that is associated with the inverse of
$\Theta^{\mu\nu}$,
\begin{eqnarray}
  \label{eq:invTheta}
  (\Theta^{-1})_{\mu\nu} = -(g - \cF)_{\mu\rho} (\cF^{-1})^{\rho\sigma}
                              (g + \cF)_{\sigma\nu} 
                            = (\cF - g \cF^{-1} g)_{\mu\nu} .
\end{eqnarray}
Therefore, associativity is obtained (even off-shell) if
\begin{eqnarray}
  \label{eq:associative}
  \bar H_{\mu\nu\rho} = 0 .
\end{eqnarray}
At this point we want to stress that we nowhere have employed the limit
$\alpha' \! \rightarrow \! 0$ in our considerations, so that the
``full'' $\Theta^{\mu\nu}$ occurs in all the relations.
This means that (\ref{eq:associative}) is a generalization of
the well known property that in the limit $\alpha' \! \rightarrow \! 0$
the product becomes associative if $H = 0$.

However, open string theory does not require such a restriction and we
investigate again the effects of the equation of motion (\ref{eq:eom}).
{}From (\ref{eq:nonassoc}) we obtain immediately that
\begin{eqnarray}
  \label{eq:intassoc}
  & & \int \ud^{D} x \sqrt {g-\cF} \; 
      \bigl( (f \circ g) \circ h - f \circ (g \circ h) \bigr) =\\
  &=& \frac {1}{6} \int \ud^{D} x \sqrt {g-\cF} \; \bigl( 
      \Theta^{\mu\sigma} \Theta^{\nu\lambda} \Theta^{\rho\kappa}
      \bar H_{\sigma\lambda\kappa} \, 
      \partial_\mu f * \partial_\nu g * \partial_\rho h \bigr) + \cO(\partial^2) \approx \nonumber\\
  &\approx& \frac {1}{6} \int \ud^{D}x \partial_\mu \bigl(\sqrt {g-\cF} 
      \ldots \bigr)^\mu + \cO(\partial^2) \; =\; 0 , \nonumber
\end{eqnarray}
so that we are allowed to omit the brackets (note that $\bar H$ is already
of order $\cO(D\Theta)$). 

For more than three functions we are allowed to omit the outermost 
bracket. In the case of four functions we obtain, for instance, the relation
\begin{eqnarray}
  \label{eq:fourfunct}
 \int \ud^{D} x \sqrt {g-\cF} \; 
      (f \circ g) \circ h \circ l =
 \int \ud^{D} x \sqrt {g-\cF} \; 
      f \circ g \circ (h \circ l) .
\end{eqnarray}
Finally, taking into account (\ref{eq:trace2}) and (\ref{eq:intassoc}),
we immediately see that the trace property holds
for an arbitrary number of functions,
\begin{eqnarray}
  \label{eq:trace}
  & & \int\!\ud^{D} x \sqrt {g-\cF} \; 
  \Bigl((\ldots( f_1 \circ \ldots))\circ f_{N-1}\Bigr) \circ f_N \approx 
\nonumber\\
  &\approx& \int\!\ud^{D} x \sqrt {g-\cF} \; 
  f_N \circ \Bigl((\ldots( f_1 \circ \ldots))\circ f_{N-1}\Bigr) \approx 
\nonumber\\
  &\approx& \int\!\ud^{D} x \sqrt {g-\cF} \; 
  \Bigl(f_N \circ (\ldots( f_1 \circ \ldots ))\Bigr)\circ f_{N-1} \approx 
  \; \ldots\ \quad .
\end{eqnarray}

We close this section with a remark on the relation to the recent work of
Cornalba and Schiappa~\cite{Cornalba:2001sm}. They considered the special 
case of a slowly varying background field $B$ in radial gauge, i.e.,
$B_{\mu\nu}(x) = B_{\mu\nu} + \frac 13 H_{\mu\nu\rho} x^\rho +\cO(x^2)$,
and a vanishing field strength $F$ for their path integral analysis.
Taking the topological limit,
$g_{\mu\nu} \sim \epsilon \rightarrow 0$,\footnote{Note 
that this limit is similar to the limit
$\alpha'\!\rightarrow\!0$ of Seiberg and Witten~\cite{Seiberg:1999vs}.}
the above properties of the product were achieved by adjusting a constant 
$\cN$ in the integration measure 
$\sqrt{B}~(1 + \cN (B^{-1})^{\mu\nu}H_{\mu\nu\rho} x^\rho)$. Using 
consistency arguments they determined the appropriate value of the 
constant to be $\cN = \frac 13$.

However, dropping the radial gauge and repeating the 
calculations\footnote{Note that in this paragraph $B_{\sigma\rho}$ denotes 
the constant part of $B_{\sigma\rho}(x)$ and all dependencies on the zero 
modes are explicitly written.}  
of~\cite{Cornalba:2001sm} for the trace property one obtains
\begin{equation}
  \label{eq:NoRadGauge}
  -i\int \sqrt{B}[\cN (B^{-1})^{\rho\sigma}(B^{-1})^{\mu\nu} 
  -(B^{-1})^{\mu\sigma}(B^{-1})^{\rho\nu}]
   \partial_\mu B_{\sigma\rho}\,f\partial_\nu g.
\end{equation}
This expression in general does not vanish for any $\cN$. Thus the trace 
property cannot be restored by an appropriate choice for the constant $\cN$ 
as it is possible for radial gauge, where $\partial_\mu B_{\sigma\rho}$ 
is replaced with $H_{\mu\sigma\rho}$.%
\footnote{It can be restored by including a term proportional to $H$ into the
measure, but such a term cannot arise from a string vacuum amplitude.}

On the other hand, expanding $B_{\mu\nu}(x)$ around its 
constant value and taking the topological limit in our setting, 
the Born-Infeld measure reduces to 
$\sqrt{B(x)}=\sqrt{B}~(1+\frac 12(B^{-1})^{\mu\nu}\partial_\rho B_{\nu\mu}x^\rho)$. 
Then (\ref{eq:NoRadGauge}) can be recast into
\begin{eqnarray}
  \label{eq:correct}
  & -i\int\sqrt{B}
   [\frac 12 (B^{-1})^{\rho\sigma}(B^{-1})^{\mu\nu}\partial_\mu B_{\sigma\rho} 
   -\frac 12 (B^{-1})^{\mu\sigma}(B^{-1})^{\nu\rho}\partial_\rho B_{\mu\sigma}&
\nonumber\\ 
  &+\frac 12 (B^{-1})^{\mu\sigma}(B^{-1})^{\nu\rho}H_{\mu\sigma\rho}]
   f\partial_\nu g 
   =\frac{i}{2}\int\sqrt{B}
    [(B^{-1})^{\mu\sigma}H_{\mu\sigma\rho}](B^{-1})^{\rho\nu}f\partial_\nu g.&
\end{eqnarray}
The last expression in square brackets in the second line is exactly what 
remains from the generalized Maxwell equation (\ref{eq:eom}) in the 
topological limit, namely the constraint 
$(B^{-1})^{\rho\sigma}H_{\rho\sigma\lambda}=0$. 
So again, the trace property holds when the background fields are on-shell! 
Nevertheless, taking the topological limit mutilates the on-shell conditions 
in the sense that no dynamics is left and only a highly restrictive nonlinear 
constraint remains. In dimensions up to four this constraint already implies 
the vanishing of the field strength $H$. Moreover, in the next order one has 
to take into account the beta function for the background metric, namely the 
Einstein equation, which imposes the even stronger restriction
\begin{eqnarray}
  \label{eq:eomlimit}
  R_{\mu\nu} - \frac 14 
      H_{\mu\rho\sigma} H^{\rho\sigma}{}_\nu 
  \sim - \frac 14 
      H_{\mu\rho\sigma} H^{\rho\sigma}{}_\nu  + 
      \cO(\epsilon^0) = 0,
\end{eqnarray}
which enforces $H_{\mu\rho\sigma}=0$ for any dimension 
(cf.~\cite{Baulieu:2001fi}).\footnote{This can be seen by first setting 
$\mu = \nu = 0$ so that $H^2_{00} = 0$ and using the antisymmetry of $H$. 
This yields $H_{0ij}=0$ for $i,j \not= 0$. The condition for purely spatial
components follows immediately.} Hence the topological limit only seems to 
make sense in the symplectic case.

\section{Conclusion}
\label{sec:concl}

On the world volume of a D-brane the product of functions (\ref{eq:2ordKonts}) 
represents a nonassociative deformation of a star product.\footnote{Note that 
in the limit of vanishing gauge fields, 
$\cF_{\mu\nu}\sim\epsilon\rightarrow 0$, the product reduces to the 
``ordinary'' product of functions and the measure reduces to $\sqrt{g}$.} 
Nevertheless, it 
enjoys the properties that the integral acts as a trace and the product of 
three functions is associative up to total derivatives. This is accomplished 
by the equations of motion of the background fields (\ref{eq:eomTheta}) and 
the Born-Infeld measure. 
No on-shell conditions have to be imposed on the inserted functions! Note, 
however, that the product of four or more functions inserted into an integral 
is ambiguous if the brackets are omitted. This is due to the fact that 
associativity for three functions is valid only up to total derivatives. 
Only the outermost lying bracket may be omitted, but this suffices 
to ensure the trace property for an arbitrary number of functions. 

Our results are complete to first order in the derivative expansion of the
background fields. 
In this approximation gravity has no influence on the generalized
product~(\ref{eq:2ordKonts}), and the sturcture is still that of the
formula given by Kontsevich.
It would be interesting to investigate whether gravity 
induces a deviation from this structure at higher orders of the derivative 
expansion. One should expect that higher order terms of the generalized 
Maxwell equation have to be used, and also the equations of motion of other
background fields may
be required to maintain the properties of the product. However, a first 
attempt in this direction showed that these calculations lead to generalized 
polylogarithms and therefore turn out to be very cumbersome.

In the near future we plan to address the question of how to use the 
open string non-commutative product and a perturbative operator product 
expansion in order to calculate correlation functions in general backgrounds. 
The property that the product of four or more functions is not unique without 
brackets seems related to the fact that these products are not independent of 
the moduli of the insertion points. For instance, in the case of four 
functions there are two distinct possibilities where to put the brackets, 
which coincides with the number of independent crossratios. This suggests 
that for 
higher $n$-point correlation functions one has to use linear combinations 
of the various orderings of the brackets weighted with coefficients 
depending on the moduli~\cite{Ho:2001qk}. 
Also the relation of these correlators to $A_\infty$ algebras \cite{stasheff}, 
the fundamental structure underlying
open-closed string field theory \cite{zwiebach}, needs further clarification.

Furthermore, it would be interesting to investigate how the 
noncommutative differential calculus is generalized in the case of a 
nonassociative algebra~\cite{Ho:2001fi}.

{\it Acknowlegements.}
A.K. was supported by the Austrian 
Research Fund FWF under grant number P14639-TPH.


\newpage

\appendix

\section{\hspace*{-20pt}ppendix: 3-point correlations and coincidence limits}
\label{sec:appx}

In the following we give an explicit calculation of the tree level 
contribution of the interaction term (\ref{eq:derFterm}), i.e., the
3-point Greens function 
$\bra\!\zeta^\mu(\tau_i) \zeta^\nu(\tau_j) \zeta^\rho(\tau_k)\!\ket$, 
which is needed in section \ref{sec:corr}. 
There we derive the correlator of two
functions $f$ and $g$. These functions contain an arbitrary power
of quantum fluctuations $\zeta^\mu$. Therefore, the correlator
has also contributions from 3-point Greens functions with two
coinciding quantum fields $\zeta^\mu$, i.e., 
$\lim_{\tau_j\rightarrow\tau_i}\bra\zeta^\mu(\tau_i)\zeta^\nu(\tau_j)\zeta^\rho(\tau_k)\ket$ or 
$\lim_{\tau_j \rightarrow \tau_k}\bra\zeta^\mu(\tau_i) \zeta^\nu(\tau_j) \zeta^\rho(\tau_k)\ket$.
The coincidence limits consist of both finite and divergent terms,
which need different treatments. The divergent ones must be
compensated by appropriate subtractions, which are accounted for in
the normal ordering of the inserted functions. The finite terms
contribute explicitly to the correlator.

We will start the appendix with the introduction of convenient
notations following~\cite{Cornalba:2001sm}. Thereafter, we derive the
Greens function
$\bra\zeta^\mu(\tau_i) \zeta^\nu(\tau_j) \zeta^\rho(\tau_k)\ket$, which
needs a regularization similar to the propagator
(\ref{eq:freepropF}). We will see that the result is a generalization
of the one in~\cite{Cornalba:2001sm}, because we do not use the
limit $\alpha'\!\rightarrow\!0$ and the radial gauge.
Finally, we perform the coincidence limits
to obtain the correct normal ordering and the finite contributions to
the correlator.

\subsection{Convenient notations and useful relations}

The free propagator (\ref{eq:fullprop}) with one side connected to the 
boundary is
\begin{eqnarray}
  \label{eq:AngleFunc}
  \bra\zeta^\mu(\tau_i)\,\zeta^\nu(z,\bar z)\ket =
  - \frac{1}{2\pi}\bigl(
  G^{\mu\nu} \cS(\tau_i,z) - \Theta^{\mu\nu} \cA(\tau_i,z)
  \bigr) ,
\end{eqnarray}
where $\cA_i$ and $\cS_i$ are defined as
\begin{eqnarray}
  \label{eq:AngleNotation}
  \cA_i =\cA(\tau_i,z)=\ln\Bigl(\frac{\bar z - \tau_i}{\bar \tau_i - z}\Bigr)
  \quad \textrm{and} \quad \cS_i = \cS(\tau_i,z) = \ln |\tau_i - \bar z|^2
  .
\end{eqnarray}
Note that $\cA_i$ is an antisymmetric function in $\tau_i$ and $z$, 
whereas $\cS_i$ is symmetric, i.e., $\cA(\tau_i,z) = -\cA(z,\tau_i)$ and
$\cS(\tau_i,z) = \cS(z,\tau_i)$. From (\ref{eq:AngleNotation}) we see that
$\cA_i$ and $\cS_i$ satisfy the relations 
$\partial \cS_i = - \partial \cA_i$ and 
$\bar\partial \cS_i = \bar\partial \cA_i$. Therefore we get
\begin{eqnarray}
  \label{eq:derivAngleFunc}
  \bra\zeta^\mu(\tau_i) \, \partial \zeta^\nu(z,\bar z)\ket &=&
    \frac{1}{2\pi} \bigl(\Theta^{\mu\nu} + G^{\mu\nu}\bigr)
    \;\partial \cA_i  \nonumber \\
  \bra\zeta^\mu(\tau_i) \, \bar\partial \zeta^\nu(z,\bar z)\ket &=&
    \frac{1}{2\pi} \bigl(\Theta^{\mu\nu} - G^{\mu\nu}\bigr)
    \;\bar\partial \cA_i .
\end{eqnarray}
Furthermore we introduce the functions
\begin{eqnarray}
  \label{eq:fAfunction}
  f_A(\tau_a,\tau_b,\tau_c) &=& 
  \int _\bbH \ud^2 z \; \partial \cA_a \bar \partial \cA_b \cA_c \\
  \label{eq:fSfunction}
  f_S(\tau_a,\tau_b,\tau_c) &=&
  \int _\bbH \ud^2 z \; \partial \cS_a \bar \partial \cS_b \cS_c =
  - \int \ud^2 z \; \partial \cA_a \bar \partial \cA_b \cS_c ,
\end{eqnarray}
which are finite except for an infinite constant. So the computation of
(\ref{eq:fAfunction}) and (\ref{eq:fSfunction}) will need a regularization.
With the above abbreviations and the relations 
\begin{eqnarray}
  \label{eq:derivGTheta}
  \partial_\rho G^{\mu\nu} &=&
     - G^{\mu\lambda} \partial_\rho\cF_{\lambda\sigma} \Theta^{\sigma\nu}
     - \Theta^{\mu\lambda} \partial_\rho\cF_{\lambda\sigma} G^{\sigma\nu}
     - \Theta^{\mu\lambda} \partial_\rho g_{\lambda\sigma} \Theta^{\sigma\nu}
     - G^{\mu\lambda} \partial_\rho g_{\lambda\sigma} G^{\sigma\nu}
     \nonumber \\
  \partial_\rho \Theta^{\mu\nu} &=&
     - \Theta^{\mu\lambda} \partial_\rho\cF_{\lambda\sigma} 
       \Theta^{\sigma\nu}
     - G^{\mu\lambda} \partial_\rho\cF_{\lambda\sigma} G^{\sigma\nu}
     - G^{\mu\lambda} \partial_\rho g_{\lambda\sigma} \Theta^{\sigma\nu}
     - \Theta^{\mu\lambda} \partial_\rho g_{\lambda\sigma} G^{\sigma\nu}
     \quad,
\end{eqnarray}
the tree level amplitude of (\ref{eq:derFterm}) reads
\begin{eqnarray}
  \label{eq:derFtree}
  \bra \zeta^{\kappa_i}(\tau_i) \, \zeta^{\kappa_j}(\tau_j) \,
      \back& &\Back \hspace{4.5pt} \zeta^{\kappa_k}(\tau_k) \; 
      \Bigl\{ - \int _\bbH \ud^2z \, 
              \partial \zeta^\mu \bar\partial \zeta^\nu \zeta^\rho \, 
              \partial_\rho (g+\cF)_{\mu\nu}
      \Bigr\} \ket_{\mathrm{tree}} \; = \nonumber \\
  = \; - \frac{1}{(2 \pi)^3} \; \Bigl\{
  \Back &+& \Back
      \Theta^{\kappa_k \rho} \partial_\rho \Theta^{\kappa_i\kappa_j}
      \;\bigl(f_A(\tau_i,\tau_j,\tau_k) - f_A(\tau_j,\tau_i,\tau_k)\bigr)
      \nonumber \\
  \Back &+& \Back
      \Theta^{\kappa_k \rho} \partial_\rho G^{\kappa_i\kappa_j}
      \;\bigl(f_A(\tau_i,\tau_j,\tau_k) + f_A(\tau_j,\tau_i,\tau_k)\bigr)
      \nonumber \\
  \Back &+& \Back
      G^{\kappa_k \rho} \partial_\rho \Theta^{\kappa_i\kappa_j}
      \;\bigl(f_S(\tau_i,\tau_j,\tau_k) - f_S(\tau_j,\tau_i,\tau_k)\bigr)
      \nonumber \\
  \Back &+& \Back
       G^{\kappa_k \rho} \partial_\rho G^{\kappa_i\kappa_j}
      \;\bigl(f_S(\tau_i,\tau_j,\tau_k) + f_S(\tau_j,\tau_i,\tau_k)\bigr)
      \nonumber \\
  \Back &+& \Back
      \bigl(\textrm{cycl. perm. (ijk)} \, \bigr) \Bigr\} .
\end{eqnarray}
For the subsequent computation of (\ref{eq:derFtree}) we take the order
$\tau_i < \tau_j < \tau_k$ on the real axis.

\subsection{Regularization of $f_A(\tau_a,\tau_b,\tau_c)$ and 
                $f_S(\tau_a,\tau_b,\tau_c)$}

To regularize $f_A$ and $f_S$ we differentiate the integral representations
(\ref{eq:fAfunction}) and (\ref{eq:fSfunction}) with respect to 
$\tau_a$, $\tau_b$ and $\tau_c$, respectively. 
Then we can perform the integration over the
upper half plane $\bbH$. This can be done by the well known method of
a transformation into a contour integral and using the residue theorem.
The pole prescriptions on the real axis are obtained 
by slightly shifting the insertion points $\tau_i$ into the upper half plane, 
so that 
\begin{eqnarray}
  \label{eq:poles}
  \cA_i = \ln \Bigl(\frac{\bar z - \tau_i - i \epsilon}
                         {\tau_i - i \epsilon - z}\Bigr)
  \quad \textrm{and} \quad 
  \cS_i = \ln \bigl( (\tau_i - i \epsilon - z)
                     (\tau_i + i \epsilon - \bar z)\bigr)
  .
\end{eqnarray}
The appearance of the logarithm needs a selection of a cut and it turns out
that the negative real axis is a convenient choice.
Finally, we determine the integrals with
respect to $\tau_a$, $\tau_b$ and $\tau_c$. Then the infinity is
contained in the integration constant. 

Thus we get
\begin{eqnarray}
  \label{eq:fresult}
  f_A(\tau_a,\tau_b,\tau_c) &=& 
  2\pi \int _0^t \ud x \Bigl( \frac{\ln(x \pm i \epsilon')}{1-x}
  + \frac{\ln(1-x \pm i \epsilon')}{x}
  \Bigr) + C^A_{(\infty)} , \nonumber\\
  f_S(\tau_a,\tau_b,\tau_c) &=&
  2\pi \int _0^t \ud x \Bigl( - \frac{\ln(x \pm i \epsilon')}{1-x}
  + \frac{\ln(1-x \pm i \epsilon')}{x} \Bigr) \\
  &-& \frac {\pi}{2} \ln^2(\tau_b - \tau_a)^2
   +  i \pi^2 \epsilon(\tau_b-\tau_a) \ln(\tau_b-\tau_a)^2 +
  C^S_{(\infty)} , \nonumber
\end{eqnarray}
where the $\pm$ in the logarithm abbreviate the
sign function $+\epsilon(\tau_b-\tau_a)$.
In (\ref{eq:fresult}) we have introduced the parameter $~t~$ which is 
defined as the combination $t = \frac {\tau_c-\tau_a}{\tau_b-\tau_a}$.
The shift $\epsilon'$ is needed to integrate along the correct side of
the cut for negative arguments of the logarithm. This selection is
determined by the pole prescription explained above.

The integrals in (\ref{eq:fresult}) lead to expressions containing
the dilogarithm which is defined as
\begin{eqnarray}
  \label{eq:dilog}
  \Li(m) = - \int _0^m \ud x \frac{\ln(1-x)}{x}
  \quad\quad \mathrm{for} \quad 0 < m < 1 .
\end{eqnarray}
with the modulus
\begin{equation}
  \label{eq:modulus}
  m = \frac {\tau_j-\tau_i}{\tau_k-\tau_i} ,
\end{equation}
which is restricted to $0 < m < 1$ because of our order
$\tau_i < \tau_j < \tau_k$.

\subsection{The tree level amplitude}

What is left is to use (\ref{eq:fresult}) to bring together all
combinations of the functions $f_A$ and $f_S$ in (\ref{eq:derFtree}).
This leads to the rather lengthy result
\begin{eqnarray}
  \label{eq:derFres}
  - 2 \pi^2
  \bra\zeta^{\kappa_i}(\tau_i)\,\Back & & \Back\zeta^{\kappa_j}(\tau_j)\,
       \zeta^{\kappa_k}(\tau_k) 
      \int _\bbH \ud^2z \, 
      \partial \zeta^\mu \bar\partial \zeta^\nu \zeta^\rho \, 
      \partial_\rho (g+\cF)_{\mu\nu}\ket_{\mathrm{tree}} \; = \nonumber 
\\
   \Bigl\{
  + \, \Theta^{\kappa_k \rho}\partial_\rho \Theta^{\kappa_i \kappa_j}
       \Back & & \Back\bigl(  \Li(1-m) - \Li(m) + \frac {\pi^2}{3}
       \bigr) \nonumber 
\\
  + \, \Theta^{\kappa_i \rho}\partial_\rho \Theta^{\kappa_j \kappa_k} 
       \Back & & \Back\bigl(  \Li(1-m) - \Li(m) - \frac {\pi^2}{3}
       \bigr)\nonumber 
\\
  +  \,\Theta^{\kappa_j \rho}\partial_\rho \Theta^{\kappa_k \kappa_i} 
       \Back & & \Back\bigl(  \Li(1-m) - \Li(m) \quad\;\;\, \bigr) \nonumber 
\\
  +  i\pi \Theta^{\kappa_k \rho}\partial_\rho G^{\kappa_i \kappa_j}
       \Back & & \Back\bigl(  \ln (m) \bigr) \nonumber 
\\
  -  i\pi \Theta^{\kappa_i \rho}\partial_\rho G^{\kappa_j \kappa_k}
       \Back & & \Back\bigl(  \ln (1-m) \bigr) \nonumber 
\\
  -  i\pi G^{\kappa_k \rho} \partial_\rho \Theta^{\kappa_i \kappa_j}
       \Back & & \Back\bigl(  \ln(\tau_k - \tau_i) \bigr)   \nonumber 
\\
  -  i\pi G^{\kappa_i \rho} \partial_\rho \Theta^{\kappa_j \kappa_k}
       \Back & & \Back\bigl(  \ln(\tau_k - \tau_i) \bigr)   \nonumber 
\\
  +  i\pi G^{\kappa_j \rho} \partial_\rho \Theta^{\kappa_k \kappa_i}
       \Back & & \Back\bigl(  \ln(\tau_k - \tau_i) \bigr)   \nonumber 
\\
  +  G^{\kappa_k \rho} \partial_\rho G^{\kappa_i \kappa_j}
     \Back & & \Back 
     \bigl(  \ln (m) \ln(1-m) - \ln^2(\tau_k - \tau_i) 
                + 2 \ln(\tau_j-\tau_i) \ln(\tau_k-\tau_i)\bigr) \nonumber 
\\
  +  G^{\kappa_i \rho} \partial_\rho G^{\kappa_j \kappa_k}
     \Back & & \Back 
     \bigl(  \ln (m) \ln(1-m) - \ln^2(\tau_k - \tau_i)
                + 2 \ln(\tau_k-\tau_j) \ln(\tau_k-\tau_i)\bigr) \nonumber 
\\
  -  G^{\kappa_j \rho} \partial_\rho G^{\kappa_k \kappa_i}
     \Back & & \Back 
     \bigl(  \ln (m) \ln(1-m) - \ln^2(\tau_k - \tau_i)\bigr)
     \quad \Bigr\} ,
\end{eqnarray}
where we have set the integration constants of (\ref{eq:fresult}) to
a convenient value, which can be done since they play no essential role
(cf. equation (\ref{eq:fullprop})).

All terms containing the boundary metric $G^{\mu\nu}$ vanish in
the limit $\alpha'\!\rightarrow\!0$. If we use, furthermore, the
radial gauge and a vanishing gauge field $A$, the terms $\pm \frac{\pi^2}3$ in
the first two lines disappear. 
Due to relation
$\Li(1-m) - \Li(m) = \frac {\pi^2}{6}(1 - 2 L(m))$, where $L(m)$ is the 
normalized Rogers dilogarithm, we thus recover the result of 
\cite{Cornalba:2001sm}.

\subsection{The coincidence limits}

In section \ref{sec:corr} we calculate the correlator of two functions.
For that purpose we have to consider the coincidence limits 
$\tau_j \rightarrow \tau_i$ and $\tau_j \rightarrow \tau_k$ of
(\ref{eq:derFres}). 

\paragraph{Singular Terms}

In these limits there appear logarithmic singularities
which can be regularized by a cut-off parameter $\Lambda$, i.e., 
$\lim _{\tau_j \rightarrow \tau_i} \ln (\tau_j - \tau_i) \rightarrow 
\ln \Lambda$. In terms of $\Lambda$ we get
\begin{eqnarray}
  \label{eq:tising}
   - \bra\zeta^{\kappa_i}(\tau_i)\,\zeta^{\kappa_j}(\tau_i)
       \Back & & \Back \zeta^{\kappa_k}(\tau_k) \;\;
       \int _\bbH \ud^2z \, 
       \partial \zeta^\mu \bar\partial \zeta^\nu \zeta^\rho \, 
       \partial_\rho (g+\cF)_{\mu\nu}\ket_{\mathrm{tree, sing}} \; = \nonumber \\  
   &=& +\frac {i}{2\pi} \Theta^{\kappa_k \rho} 
                     \partial_\rho G^{\kappa_i \kappa_j}
                     \ln \Lambda
       +\frac{1}{\pi^2} G^{\kappa_k \rho} 
                     \partial_\rho G^{\kappa_i \kappa_j}
                     \ln \Lambda \ln (\tau_k - \tau_i) \nonumber \\
   &=& - \frac {1}{\pi} \partial_\rho G^{\kappa_i \kappa_j} 
                               \ln \Lambda
       \bra\zeta^\rho(\tau_i)\,\zeta^{\kappa_k}(\tau_k)\ket
\end{eqnarray}
for $\tau_j \rightarrow \tau_i$ and 
\begin{eqnarray}
  \label{eq:tksing}
   - \bra\zeta^{\kappa_i}(\tau_i)\Back & & \Back
      \zeta^{\kappa_j}(\tau_k)\,\zeta^{\kappa_k}(\tau_k)
      \int _\bbH \ud^2z \, 
      \partial \zeta^\mu \bar\partial \zeta^\nu \zeta^\rho \, 
      \partial_\rho (g+\cF)_{\mu\nu}\ket_{\mathrm{tree, sing}} \; = \nonumber \\  
   &=& -\frac {i}{2\pi} \Theta^{\kappa_i \rho} 
                     \partial_\rho G^{\kappa_j \kappa_k}
                     \ln \Lambda
       +\frac{1}{\pi^2} G^{\kappa_i \rho} 
                     \partial_\rho G^{\kappa_j \kappa_k}
                     \ln \Lambda \ln (\tau_k - \tau_i)\nonumber \\
   &=&  - \frac {1}{\pi} \partial_\rho G^{\kappa_j \kappa_k} 
                               \ln \Lambda
       \bra\zeta^{\kappa_i}(\tau_i)\,\zeta^\rho(\tau_k)\ket      
\end{eqnarray}
for $\tau_j \rightarrow \tau_k$.
The singularities (\ref{eq:tising}) and (\ref{eq:tksing})
must be compensated by
appropriate counter terms.
The correct subtractions 
can easily be read off from (\ref{eq:tising},\ref{eq:tksing}). 
Together with the singular part of the propagator (\ref{eq:boundaryprop}) 
we get
\begin{eqnarray}
  \label{eq:ope}
  \zeta^\mu (\tau) \, \zeta^\nu (\tau')  &=& 
      - \frac {1}{2\pi} G^{\mu\nu} \ln(\tau-\tau')^2
      - \frac {1}{2\pi} 
        \partial_\rho G^{\mu \nu} \ln(\tau-\tau')^2 \; 
        \zeta^\rho (\frac {\tau+\tau'}{2}) \nonumber \\
  &+& ( \; \textrm{regular terms} \; ) .
\end{eqnarray}

\paragraph{Finite Terms}

In our limits equation (\ref{eq:derFres}) contains also finite parts which read
\begin{eqnarray}
  \label{eq:tifinite}
   - \Back &\bra&\Back\zeta^{(\kappa_i}(\tau_i)\,\zeta^{\kappa_j)}(\tau_i) \;\;
       \zeta^{\kappa_k}(\tau_k) \;\;
      \int _\bbH \ud^2z \, 
      \partial \zeta^\mu \bar\partial \zeta^\nu \zeta^\rho \, 
      \partial_\rho (g+\cF)_{\mu\nu}\ket_{\mathrm{tree, fin}} \; = \nonumber \\
   &=& - \frac {1}{12} \bigl(
         \Theta^{\kappa_i \rho} \partial_\rho \Theta^{\kappa_j \kappa_k} -
         \Theta^{\kappa_j \rho} \partial_\rho \Theta^{\kappa_k \kappa_i}
         \bigr) \nonumber \\
   & & - \frac {i}{2\pi} \bigl(
         \Theta^{\kappa_k \rho} \partial_\rho G^{\kappa_i \kappa_j} +
         G^{\kappa_j \rho} \partial_\rho \Theta^{\kappa_i \kappa_k} +
         G^{\kappa_i \rho} \partial_\rho \Theta^{\kappa_j \kappa_k} \bigr) 
         \ln(\tau_k-\tau_i) \nonumber\\
   & & - \frac {1}{2\pi^2} \bigl(
         G^{\kappa_k \rho} \partial_\rho G^{\kappa_i \kappa_j} -
         G^{\kappa_i \rho} \partial_\rho G^{\kappa_j \kappa_k} -
         G^{\kappa_j \rho} \partial_\rho G^{\kappa_k \kappa_i} \bigr)
         \ln^2 (\tau_k-\tau_i)
\end{eqnarray}
for $\tau_j \rightarrow \tau_i$ and 
\begin{eqnarray}
  \label{eq:tkfinite}
   - \Back&\bra&\Back \zeta^{\kappa_i}(\tau_i) \;\; 
      \zeta^{(\kappa_j}(\tau_k) \, \zeta^{\kappa_k)}(\tau_k) \;\;
      \int _\bbH \ud^2z \, 
      \partial \zeta^\mu \bar\partial \zeta^\nu \zeta^\rho \, 
      \partial_\rho (g+\cF)_{\mu\nu}\ket_{\mathrm{tree, fin}} \; = \nonumber \\  
   &=& + \frac {1}{12} \bigl(
         \Theta^{\kappa_k \rho} \partial_\rho \Theta^{\kappa_i\kappa_j} -
         \Theta^{\kappa_j \rho} \partial_\rho \Theta^{\kappa_k\kappa_i}
         \bigr) \nonumber \\
   & & + \frac {i}{2\pi} \bigl(
         \Theta^{\kappa_i \rho} \partial_\rho G^{\kappa_j \kappa_k} +
         G^{\kappa_j \rho} \partial_\rho \Theta^{\kappa_k \kappa_i} +
         G^{\kappa_k \rho} \partial_\rho \Theta^{\kappa_j \kappa_i} \bigr) 
         \ln(\tau_k-\tau_i) \nonumber\\
   & & - \frac {1}{2\pi^2} \bigl(
         G^{\kappa_i \rho} \partial_\rho G^{\kappa_j \kappa_k} -
         G^{\kappa_j \rho} \partial_\rho G^{\kappa_k \kappa_i} -
         G^{\kappa_k \rho} \partial_\rho G^{\kappa_i \kappa_j} \bigr)
         \ln^2 (\tau_k-\tau_i)
\end{eqnarray}
for $\tau_j \rightarrow \tau_k$. We have taken into account only the
symmetric part of the limit, since the antisymmetric one does not
contribute in section \ref{sec:corr}.

\end{document}